\documentclass[12pt]{article}

\usepackage[dvips]{graphicx}

\usepackage{amsmath}
\newcommand{\lsim}{\raisebox{-4pt}{$\,\stackrel{\textstyle
                                                         <}{\sim}\,$}}

\newcommand{\nn}{\nonumber}
\newcommand{\be}{\begin{equation}}
\newcommand{\ee}{\end{equation}}
\newcommand{\ba}{\begin{eqnarray}}
\newcommand{\ea}{\end{eqnarray}}
\newcommand{\req}[1]{(\ref{#1})}
\def\={\,=\,}
\newcommand{\ci}[1]{\cite{#1}}

\def\mev{\operatorname{MeV}}
\def\gev{\operatorname{GeV}}

\def\ale{\alpha_{\rm em}}
\def\als{\alpha_{\rm s}}

\def\LQCD{\Lambda_{\rm QCD}}

\newcommand{\tw}{\textwidth}
\def\vk{{\bf k}_{\perp}}

\def\vbs{{\bf b}}
\def\vb0{{\bf b}_0}

\newcommand{\Sla}{\hspace*{-0.020\tw}/}

\def\={\,=\,}

\begin{document} 
\thispagestyle{empty}
\begin{flushright}
WU B 15-03 \\
June, 10  2015\\[20mm]
\end{flushright}

\begin{center}
{\Large\bf The exclusive limit of the pion-induced Drell-Yan process}\\
\vskip 10mm

S.V.\ Goloskokov
\footnote{Email:  goloskkv@theor.jinr.ru}
\\[1em]
{\small {\it Bogoliubov Laboratory of Theoretical Physics, Joint Institute
for Nuclear Research, Dubna 141980, Moscow region, Russia}}\\

\vskip 5mm
P.\ Kroll \footnote{Email:  kroll@physik.uni-wuppertal.de}
\\[1em]
{\small {\it Fachbereich Physik, Universit\"at Wuppertal, D-42097 Wuppertal,
Germany}}\\
and
{\small {\it Institut f\"ur Theoretische Physik, Universit\"at
    Regensburg, \\D-93040 Regensburg, Germany}}\\

\end{center}
\vskip 5mm 
\begin{abstract}
\noindent 
Based on previous studies of hard exclusive leptoproduction of pions
in which the essential role of the pion pole and the transversity generalized
parton distributions (GPDs) has been pointed out, we present predictions
for the four partial cross sections of the exclusive Drell-Yan process,
$\pi^-p\to l^-l^+n$.
\end{abstract}   

\section{Introduction}
In recent years hard exclusive leptoproduction of mesons and photons have been
studied intensively by both experimentalists and theoreticians. It became evident
in the course of time that within the handbag approach which is based on QCD 
factorization in the generalized Bjorken regime of large photon virtuality 
and large photon-proton center-of-mass energy but fixed $x$-Bjorken, it is 
possible to interpret these processes in terms of generalized parton distributions  
and hard perturbatively calculable subprocesses with, however, occasionally strong 
power corrections for meson production (for a recent review see \ci{kroll14}).
Exploiting the universality property of the GPDs, one may use the set of GPDs 
extracted from meson leptoproduction, in the calculation of other hard exclusive 
processes. Of particular interest are processes with time-like virtual photons.
Thus in \ci{pire13} predictions for time-like DVCS ($\gamma p\to l^-l^+p$) have been 
given, their experimental examination is still pending. The high-energy pion
beam at J-PARC put into operation in the near future, offers the possibility of 
measuring another exclusive process with time-like virtual photons, namely the 
exclusive limit of the Drell-Yan process, $\pi^-p\to l^-l^+n$. The purpose of this 
letter is to present predictions for the cross sections of this process taking into 
account what has been learned in the analyses of pion leptoproduction \ci{GK5,GK6}. 
The data on the cross section for $\pi^+$ leptoproduction \ci{hermes-cs-pip,F-pi} 
demonstrate the prominent role of the contribution from the pion pole at small
invariant momentum transfer, $t$, and it became evident that it is to be calculated 
as an one-particle-exchange (OPE) term rather than from the GPD $\widetilde E$ 
\ci{mankiewicz98}. In the latter case the pion-pole 
contribution to the $\pi^+$ cross section is underestimated by order of magnitude. 
A second important observation has been made in \ci{GK5,GK6}: The interpretation of the 
transverse target spin asymmetries in $\pi^+$ leptoproduction measured by the 
HERMES collaboration \ci{hermes-aut} necessitates contributions from transversely 
polarized photons which are to be modelled by transversity GPDs within the handbag 
approach. This observation is supported by a recent CLAS measurement of $\pi^0$ 
leptoproduction \ci{clas-pi0}. 

Since for the process $\pi^-p\to l^-l^+n$. the same GPDs contribute as for pion 
leptoproduction and the corresponding subprocesses are just 
$\hat s \leftrightarrow \hat u$ crossed ones~\footnote{
A detailed discussion of the space- and time-like connection of the leading-twist
amplitudes can be found in \ci{mueller12}}
\be
{\cal H}^{\pi^-\to\gamma^*}(\hat s, \hat u)\=-{\cal H}^{\gamma^*\to\pi^+}(\hat u,\hat s)
\ee
where $\hat s$ and $\hat u$ denote the subprocess Mandelstam variables, one can 
exploit the knowledge acquired there. One thus gains predictive power, there is no 
free parameter or soft hadronic matrix element left for the Drell-Yan process. 
Our analysis markedly differs from a previous study performed by Berger {\it et al} 
\ci{berger01} where only predictions for the longitudinal cross section at 
leading-twist accuracy has been given. It should be stressed that their and our 
predictions for that cross section differ by about a factor of 40 due to the 
different treatment of the pion pole contribution. Our findings may be of help in 
the preparation of an Drell-Yan experiment \ci{chang}. Future data on 
the exclusive pion-induced Drell-Yan process may reveal whether or not our present 
understanding of hard exclusive processes in terms of convolutions of GPDs and hard 
subprocesses also holds for time-like photons . This is a non-trivial issue because 
the physics in the time-like region is complicated and often not understood. Thus, 
for instance, there is no explanation of the time-like electromagnetic form factors
of hadrons \ci{bakulev00}. Even the semi-inclusive Drell-Yan process was difficult 
to understand. It took a long time before the discrepancy between the theoretical 
predictions and experiment, known as the $K$-factor, has been explained as 
threshold logarithms \ci{sterman87,catani89} representing gluon radiation resummed 
to next-leading-log (NLL) accuracy. 

\section{The handbag approach}
\begin{figure}[t]
\centering
\includegraphics[width=0.48\tw]{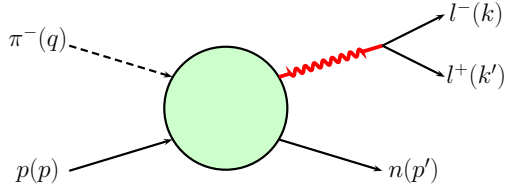}
\caption{The exclusive Drell-Yan process. The symbols in brackets denote the momenta
of the respective particles.}
\label{fig:process}
\end{figure}

Here, in this section, we recapitulate the handbag approach. 
For more details of it we refer to our previous work \ci{GK5,GK6}.
The process $\pi^-p\to l^-l^+n$ is depicted in Fig.\ \ref{fig:process}.
We work in a center-of-mass frame in which ${\bf p}+{\bf p}'$ points along the 
positive 3-axis and we consider the kinematical range of large Mandelstam $s$ 
($=(p+q)^2$) and large photon virtuality~\footnote{
The $Q^{\prime 2}$-regions of quarkonia states have to be excluded.}
, $Q^{\prime 2}$, but small 
\be
\tau\=\frac{Q^{\prime 2}}{s-m^2}\,,
\ee
the time-like analogue of Bjorken-$x$ ($m$ being the mass of the nucleon). 
Hence, skewness, defined as
\be
\xi\=\frac{p^+-p^{\prime +}}{p^+ +p^{\prime +}}\approx \frac{\tau}{2-\tau}\,,
\ee
is also small. 

Assuming factorization we can express the helicity amplitudes for $\pi^-p\to\gamma^*n$ 
in terms of convolutions of GPDs and hard subprocess amplitudes 
\ba
{\cal M}_{0+,0+} &=& \sqrt{1-\xi^2}\, \frac{e_0}{Q^\prime}
                             \,\Big[\langle \widetilde{H}^{(3)}\rangle
          -\frac{\xi^2}{1-\xi^2}\langle \widetilde{E}^{(3)}_{\rm n.p.}\rangle 
          + \frac{2\xi m}{1-\xi^2}\frac{\varrho_\pi}{t-m_\pi^2}\Big]\,,\nn\\
{\cal M}_{0-,0+} &=& \frac{\sqrt{-t^\prime}}{2m}\,\frac{e_0}{Q^\prime}\,
             \Big[ \xi \langle \widetilde{E}^{(3)}_{\rm n.p.}\rangle 
              - 2m\frac{\varrho_\pi}{t-m_\pi^2}\Big]\,, \nn\\
{\cal M}_{--,0+} &=& \sqrt{1-\xi^2}\, \frac{e_0}{Q^{\prime 2}}\, 
                                      \mu_\pi \langle H_T^{(3)}\rangle\,,\nn\\
{\cal M}_{\pm+,0+}&=&\frac{\sqrt{-t^\prime}}{4m}\,\frac{e_0}{Q^{\prime 2}}\,
                \Big[\mu_\pi \langle \bar{E}_T^{(3)}\rangle \mp 8\sqrt{2}m^2\xi
                          \frac{\varrho_\pi}{t-m_\pi^2}\Big]\,, \nn\\
{\cal M}_{+-,0+}&\approx& 0\,.
\label{eq:amplitudes}
\ea
Explicit helicities are labeled by their signs or by zero, $e_0$ denotes the 
positron charge and $t'=t-t_0$ where $t_0=-4m^2\xi^2/(1-\xi^2)$ is the minimal 
value $t$ corresponding to forward scattering. Terms of order $t/Q^{\prime 2}$ are 
neglected throughout. The amplitudes for negative helicity of the initial state 
proton are obtained from the set of amplitudes \req{eq:amplitudes} by parity 
conservation. The residue of the pion pole is given by
\be
\varrho_\pi\=\sqrt{2} g_{\pi NN} F_{\pi NN}(t) Q^{\prime 2}F_\pi(Q^{\prime 2})
\label{eq:rho-pi}
\ee
where $g_{\pi NN}$ ($=13.1\pm 0.3$) is the familiar pion-nucleon coupling constant 
and $F_{\pi NN}$ is a form factor that describes the $t$-dependence of the coupling 
of the virtual pion to the nucleon. The pion mass, $m_\pi$, is neglected except 
in the pion propagator. As we mentioned in the introduction we treat the pion 
pole as an OPE term. Therefore the full time-like electromagnetic form factor 
occurs in \req{eq:rho-pi}. Calculating the pion pole contribution from the GPD 
$\widetilde E$ as it is done in \ci{berger01}, one obtains to same the expression  
for it but with the leading-order (LO) perturbative result for the pion form factor. 
In \req{eq:amplitudes} it is also allowed for a possible non-pole (n.p.) part 
of $\widetilde E$. 

For incident $\pi^-$ mesons the $p\to n$ transition GPDs are required which, as a
consequence of isospin invariance, are given by the isovector combination of proton
GPDs \ci{mankiewicz98}
\be
K^{(3)}\=K^u-K^d\,.
\label{eq:isovector}
\ee
The convolutions of the GPDs and the amplitudes ${\cal H}$ for the subprocess 
$\pi^-q\to\gamma^*q$ read \ci{GK5,GK6}
\be
\langle K^{(3)} \rangle \=\int dx {\cal H}_{\mu\lambda,0+}(x,\xi,Q^{\prime 2},t\simeq 0)
                                 K^{(3)}(x,\xi,t)\,.
\ee
The helicity of the final state quark is $\lambda=\mu +1/2$ with the photon helicity, 
$\mu$, being either zero or -1. Thus, the asymptotically leading longitudinal amplitude 
is related to a helicity-non-flip subprocess amplitude while, for transverse photons, 
a helicity-flip amplitude is convoluted with the transversity GPDs $H_T$ and the 
combination $\bar{E}_T=2\widetilde{H}_T+E_T$. As made explicit in \req{eq:amplitudes} 
the transverse amplitudes are suppressed by $\mu_\pi/Q^{\prime}$ as compared to the 
longitudinal ones. The mass parameter $\mu_\pi$ is related to the chiral condensate
\be
\mu_\pi\=\frac{m_\pi^2}{m_u+m_d}
\label{eq:condensate}
\ee 
($m_u, m_d$ are current quark masses). The subprocess amplitudes are calculated to LO 
of perturbation theory retaining quark transverse momenta, $\vk$, and taking into 
account Sudakov suppressions while the emission and reabsorption of partons by the 
nucleon happens collinearly to the nucleon momenta. This so-called modified 
perturbative approach turns into the leading-twist result \ci{berger01} for 
$Q^{\prime 2}\to \infty$. 

Since the Sudakov factor, $\exp[-S]$, comprises gluonic radiation, resummed to all 
orders of perturbation theory in NLL approximation \ci{li-sterman} which can only be 
efficiently performed in the impact parameter space, canonically conjugated to the 
$\vk$-space, one is forced to work in the $\vbs$-space. Hence,
\ba
{\cal H}_{\mu\lambda,0+}&=&\int dz d^2b\, \hat{\Psi}_{-\lambda +}(z,-\vbs) 
          \hat{F}_{\mu\lambda,0 +}(x,\xi,z,Q^{\prime 2},\vbs) \nn\\
           &\times& \als(\mu_R) \exp{[-S(z,\vbs,Q^{\prime 2})]}\,.
\ea
The Fourier transforms of the hard scattering kernel and the light-cone wave function 
of the pion are denoted by $\hat F$ and $\hat \Psi$, respectively. The momentum 
fraction of the helicity $+1/2$ quark entering the pion is denoted by $z$; the 
helicity of the antiquark is $-\lambda$. For the renormalization scale we choose 
$\mu_R={\rm max}(z Q^\prime, (1-z)Q^\prime,1/b)$ and the factorization scale is $1/b$.
Following Li and Sterman \ci{li-sterman} we only retain the most important quark 
transverse momenta which appear in the denominators of the parton propagators in the 
hard scattering kernels. Therefore, we can use the light-cone projector of a $q\bar{q}$ 
pair on an ingoing pion in collinear approximation \ci{beneke}
\ba
{\cal P}_\pi&=&\frac{f_\pi}{2\sqrt{2N_c}}\,\frac{\gamma_5}{\sqrt{2}} 
                        \left\{q\,\Sla\; 
                \Phi(z) \phantom{\frac{\partial}{\partial}} \right. \nn\\
            &&\left. -\mu_\pi \Big[\Phi_P(z) 
       -i\frac{\sigma_{\mu\nu}}{2N_c}\Big(\frac{q^\mu n^\nu}{q\cdot n}\Phi_\sigma(z)
        -q^\mu\frac{d\Phi_\sigma(z)}{dz} \frac{\partial}{\partial k_{\perp \nu}}
                  \Big)\Big]\right\}\,.
\label{eq:projector}
\ea
and replace the distribution amplitudes by light-cone wave functions. In \req{eq:projector}
$f_\pi (=132\,\mev)$ is the pion decay constant, $N_c$ the number of colors and 
$n$ is a light-like vector which in a frame where the massless pion moves along 
the $z$-direction is $n=[0,1,0_\perp]$. Three-particle configurations, $q\bar{q}g$, 
are neglected. Dirac, flavor and color labels are omitted for convenience. The 
first term in \req{eq:projector} is the well-known twist-2 part which is employed 
in the calculation of ${\cal H}_{0\pm,0+}$. For the accompanying light-cone wave function 
we take
\be
\Psi_{-+}\=\frac{\sqrt{2N_c}}{f_\pi} \exp{[-a^2_\pi \vk^2/(z(1-z))]}
\ee
with the transverse size parameter $a_\pi=\big[\sqrt{8}\pi f_\pi\big]^{-1}$ fixed 
from $\pi^0\to\gamma\gamma$ decay \ci{huang}.
The twist-3 part of \req{eq:projector} is utilized in the calculation of 
${\cal H}_{--,0+}$. As the calculation reveals this subprocess amplitude
is dominated by the contribution from $\Phi_P$ while the tensor term provides
a correction of order $t/Q^{\prime 2}$ which is neglected for consistency. For the
wave function associated to $\Phi_P$ ($\equiv 1$), we use~\footnote
{Since quark and antiquark forming the pion have the same helicity, it may seem 
appropriate to use a $l_z=\pm 1$ wave function (for a particle moving along the 
$z$-direction). Such a wave function has been proposed in \ci{yuan-ji}. It is 
proportional to $k_\perp^{\pm}=k_\perp^x\pm i k_\perp^y$. Its collinear reduction leads 
to the tensor piece in \req{eq:projector}, the important term $\sim \Phi_P$ is 
lacking.}
\be
\Psi_{++}\=\frac{16\pi^{3/2}}{\sqrt{2N_c}} f_\pi a_P^3|\vk| 
                  \exp{[-a_P^2\vk^2]}\,.
\ee
For the transverse size parameter, $a_P$, we take $1.8\,\gev^{-1}$. 

\section{Predictions for the partial cross sections}
Before we present our predictions for the exclusive Drell-Yan process
we specify the various parameters and soft hadronic functions we use in the 
evaluation. The form factor $F_{\pi NN}$ appearing in \req{eq:rho-pi}, is 
parametrized as
\be
F_{\pi NN}\=\frac{\Lambda_N^2-m_\pi^2}{\Lambda_N^2\,-\,t\;}
\ee
with $\Lambda_N=0.44\pm 0.04\,\gev$. 
For the time-like pion electromagnetic form factor also occuring in 
\req{eq:rho-pi}, we take the average of the data from CLEO \ci{cleo} and BaBar 
\ci{babar} as well as a value derived from the $J/\Psi\to \pi^+\pi^-$ decay \ci{PDG}
\be
Q^{\prime 2} |F_\pi(Q^{\prime 2})| \= 0.88\pm 0.04\,\gev^2\,.
\label{eq:pionFF}
\ee
For its phase, $\exp{[i\delta(Q^{\prime 2})]}$, we rely on a recent dispersion 
analysis \ci{belicka} which, for $2\,\gev^2 \lsim Q^{\prime 2} \lsim 5\,\gev^2$, 
provides
\be
\delta\=1.014 \pi + 0.195(Q^{\prime 2}/\hspace*{-0.005\tw}\gev^2 -2)  
          - 0.029(Q^{\prime 2}/\hspace*{-0.005\tw}\gev^2 -2)^2\,.
\ee
In the absence of any other information on this phase we use this parametrization
up to $\approx 8.9\,\gev^2$ where $\delta=\pi$. For larger values of $Q^{\prime 2}$
we take $\delta=\pi$, the asymptotic phase of the time-like pion form factor 
obtained by analytic continuation of the perturbative result for the space-like 
form factor \ci{bakulev00}.   

The GPDs are constructed with the help of the familiar double distribution ansatz 
from the zero-skewness GPDs which are parametrized as~\footnote
{A more complicated profile function is adopted for $\widetilde H$, see \ci{DK13}.}
\be
K(x,\xi=0,t)\=k(x) \exp{[t(b+\alpha'\ln{x}]}
\ee
where the forward limit, $k(x)$, is an appropriate parton distribution (PDF) or 
is parametrized like a PDF with parameters fitted to experiment. The GPD
$\widetilde H (x,\xi=0,t)$ (including an error estimate) is taken from the recent 
analysis of the nucleon form factors \ci{DK13} which, for this GPD, is based 
on the DSSV polarized PDFs \ci{dssv09}. A non-pole contribution to $\widetilde E$ 
is neglected, there is no clear signal for it in the data on pion leptoproduction. 
For the zero-skewness transversity GPDs, $H_T$ and $\bar{E}_T$, the actual values 
of the parameters are specified in \ci{GK7}. They are oriented on  lattice QCD 
results \ci{QCDSF05,QCDSF06} and lead to fair fits of the pion leptoproduction data 
\ci{hermes-cs-pip,hermes-aut,clas-pi0} as well as of the spin density matrix elements 
and transverse target spin asymmetries for vector mesons \ci{GK8,GK7}. The errors of 
the transversity GPDs are estimated from the $p$-pole fits presented in 
\ci{QCDSF05,QCDSF06}. For the mass parameter \req{eq:condensate} that controls the 
strength of the twist-3 amplitudes, we adopt the value~\footnote{
According to the recent particle data tables \ci{PDG} $\mu_\pi$ is rather $2.6\,\gev$.
Using this value the normalizations of $H_T$ and $\bar{E}_T$ have to be altered 
accordingly since the fit to the pion leptoproduction data fixes the product of 
$\mu_\pi$ and the transversity GPDs.}
$\mu_\pi=2\,\gev$ valid at the scale $2\,\gev$. For its error we choose $+0.55$ and 
$-0.15$ \ci{PDG}. The QCD coupling constant, $\als$, is evaluated from the 
one-loop epression for four flavors and  $\Lambda_{\rm QCD}=182\,\mev$. The 
time-like Sudakov factor is unknown, the continuation from the space-like to the 
time-like region is not well understood (see \ci{gousset-pire}). The replacement of 
$Q^2$ by $-Q^{\,\prime 2}$ (see \ci{gousset-pire,magnea}) leads to an oscillating phase 
but it is unclear whether these oscillations are physical or not. We therefore follow
Gousset and Pire \ci{gousset-pire} and use the space-like Sudakov factor, as utilized 
in our previous work, also in the time-like region (with $Q^2\to Q^{\,\prime 2}$). As 
shown in \ci{jakob-kroll}, for $Q^{\prime 2}$ less than $10\,\gev^2$ the Sudakov fator 
is always close to unity except near $b=1/\LQCD$ where it drops to zero sharply. With 
the exception of this region the wave function provides the main suppression. 
Therefore, the detailed behavior of the Sudakov factor is not very important.

\begin{figure}
\centering
\includegraphics[width=0.48\tw]{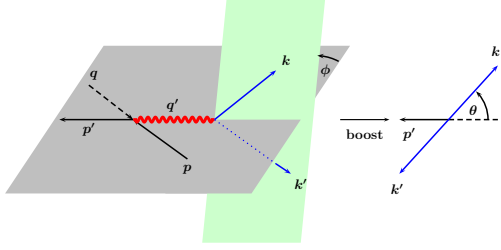}
\caption{Definition of the angles $\phi$ and $\theta$. The latter angle is defined
in the rest frame of the virtual photon.}
\label{fig:frames}
\end{figure}
The four-fold differential cross section for $\pi^-p\to l^-l^+n$ read
\ba
\frac{d\sigma}{dt dQ^{\prime 2} d\cos{\theta}d\phi} &=&\frac{3}{8\pi}\left\{
            \sin^2{\theta}\,\frac{d\sigma_L}{dtdQ^{\prime 2}}
+\frac{1+\cos^2{\theta}}{2}\, \frac{d\sigma_T}{dtdQ^{\prime 2}} \right.\nn\\
 &+&\left.\frac{\sin{(2\theta)}\cos{\phi}}{\sqrt{2}} \,
               \frac{d\sigma_{LT}}{dtdQ^{\prime 2}}
     + \sin^2{\theta}\cos{(2\phi)}\,\frac{d\sigma_{TT}}{dtdQ^{\prime 2}}\right\}
\ea
where the angles $\phi$ and $\theta$, specifying the directions of the leptons, are 
defined in Fig.\ \ref{fig:frames}. The partial cross sections are related to the 
$\pi^-p\to \gamma^*n$ helicity amplitudes \req{eq:amplitudes} by
\ba
\frac{d\sigma_L}{dt dQ^{\prime 2}}&=& \kappa\, \sum_{\nu'}|{\cal M}_{0\nu',0+}|^2\,,\nn\\
\frac{d\sigma_T}{dt dQ^{\prime 2}}&=& \kappa\, \sum_{\mu=\pm1,\nu'}
                         |{\cal M}_{\mu\nu',0+}|^2\,,\nn\\
\frac{d\sigma_{LT}}{dtdQ^{\prime 2}}&=&\kappa\,\operatorname{Re}\sum_{\nu'}
                             \big[{\cal M}^*_{0\nu',0+}
                          ({\cal M}_{+\nu',0+}-{\cal M}_{-\nu',0+})\big]\,,\nn\\
\frac{d\sigma_{TT}}{dtdQ^{\prime 2}}&=&\kappa\,\operatorname{Re}\sum_{\nu'}
           \big[{\cal M}^*_{+\nu',0+}{\cal M}_{-\nu',0+}\big].
\label{eq:partial}
\ea
The normalization factor reads (lepton masses are neglected)
\be
\kappa \= \frac{\ale}{48\pi^2}\frac{1}{(s-m^2)^2 Q^{\prime 2}}\,.
\ee

\begin{figure}[th]
\centering
\includegraphics[width=0.35\tw]{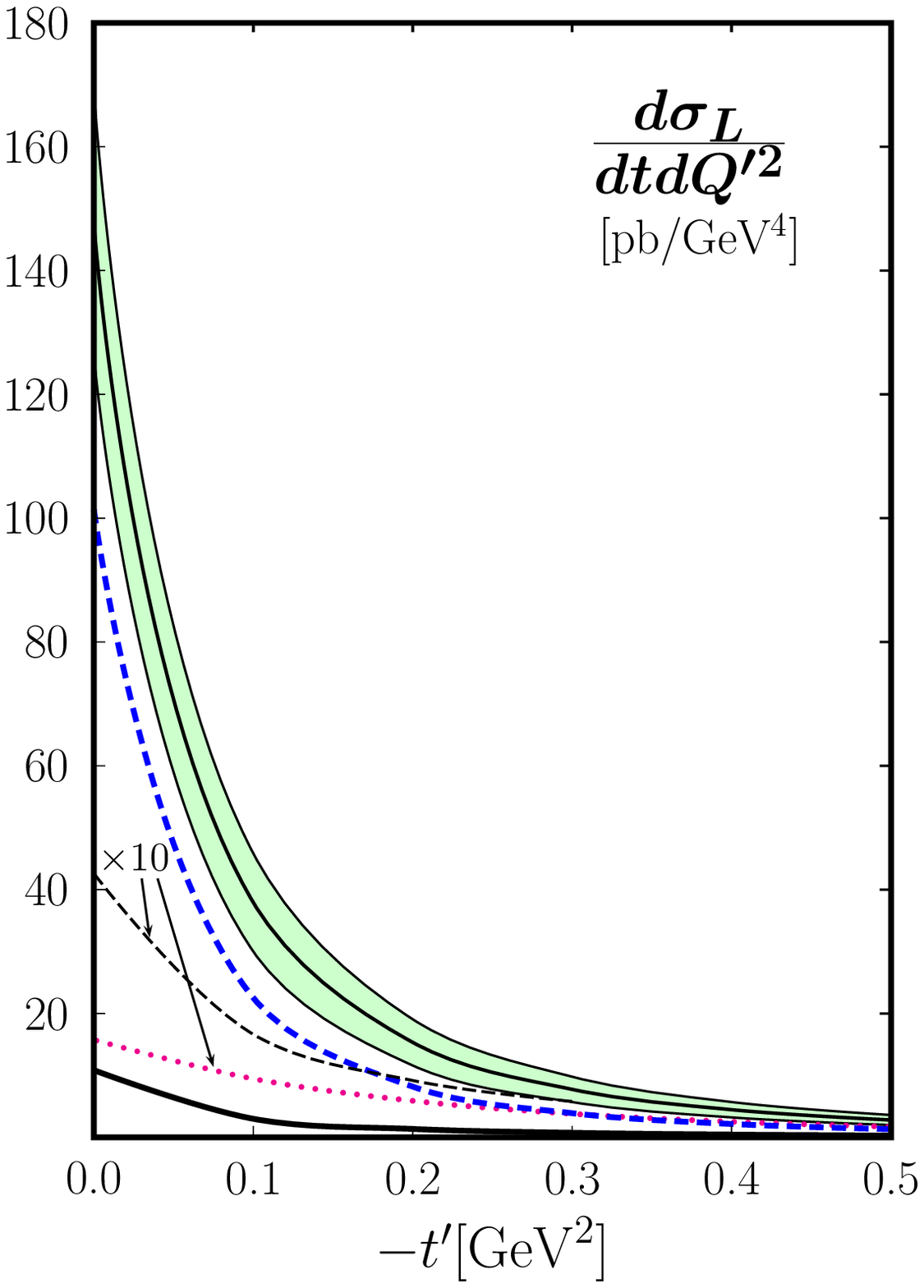}
\hspace*{0.06\tw}
\includegraphics[width=0.345\tw]{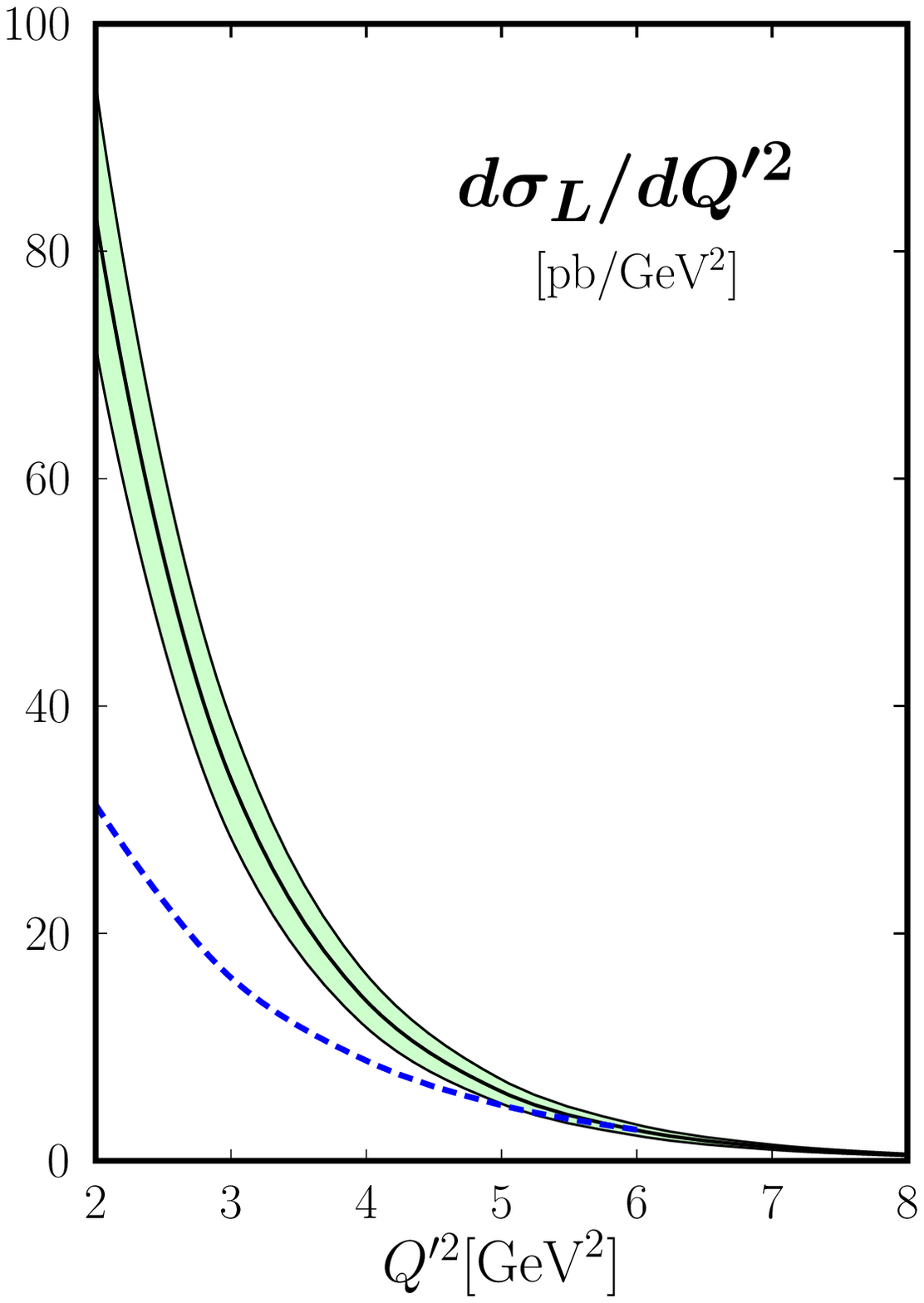}
\caption{The longitudinal cross sections $d\sigma_L/dtdQ^{\prime 2}$ (left) at 
$Q^{\prime 2}=4\,\gev^2$ versus $t'$  and $d\sigma_L/dQ^{\prime 2}$ (right) versus 
$Q^{\prime 2}$. The thin solid lines with error bands represent our full results 
at $s=20\,\gev^2$, the thick dashed ones those at $30\,\gev^2$. The thick solid
(dotted, thin dashed) line is the interference term (contribution from 
$|\langle \widetilde{H}^{(3)}\rangle|^2$, leading twist). The latter two results 
are multiplied by 10 for the ease of legibility.}
\label{fig:sigmaL}
\end{figure}
In Fig.\ \ref{fig:sigmaL} we show our predictions for $d\sigma_L/dtdQ^{\prime 2}$    
at $Q^{\prime 2}=4\,\gev^2$ and $s=20\,\gev^2$ and $d\sigma_L/dQ^{\prime 2}$     
integrated over $t'$ from 0 to $-0.5\,\gev^2$. The longitudinal cross section is 
heavily dominated by the contribution from the pion pole, that one from 
$\widetilde H$, including its interference with the pion pole, amounts only to 
about $10\%$ in the kinematical range of interest. The full result is markedly 
larger than our leading-twist result which is of the same order as that one 
quoted in \ci{berger01}. This amplification is due to the use of the experimental 
value of the pion form factor \req{eq:pionFF} instead of its leading-twist result 
($\approx 0.15\,\gev^2$). We stress that the OPE contribution from the pion pole 
does neither rely on QCD factorization nor on a hard scattering. It is therefore 
not subject to evolution and higher-order perturbative QCD corrections. Because of 
the dominant contribution from the pion-pole and since we only consider a small 
range of $Q^{\prime 2}$ around $4\,\gev^2$ the evolution of the GPDs is insignificant 
and is therefore neglected. As opposed to \ci{berger01} our interference term is 
positive. It is generated by the imaginary parts of $\langle {\widetilde H}\rangle$ 
and the pion-pole contribution while, in a LO leading-twist calculation, it is 
evidently under control of the corresponding real parts. Constructing $\widetilde H$ 
from the polarized PDFs derived in \ci{BB} instead from the DSSV ones \ci{dssv09}  
alters the predictions for the longitudinal cross section by less than the estimated 
errors displayed in Fig.\ \ref{fig:sigmaL}. 

\begin{figure}[t]
\centering
\includegraphics[width=0.35\tw]{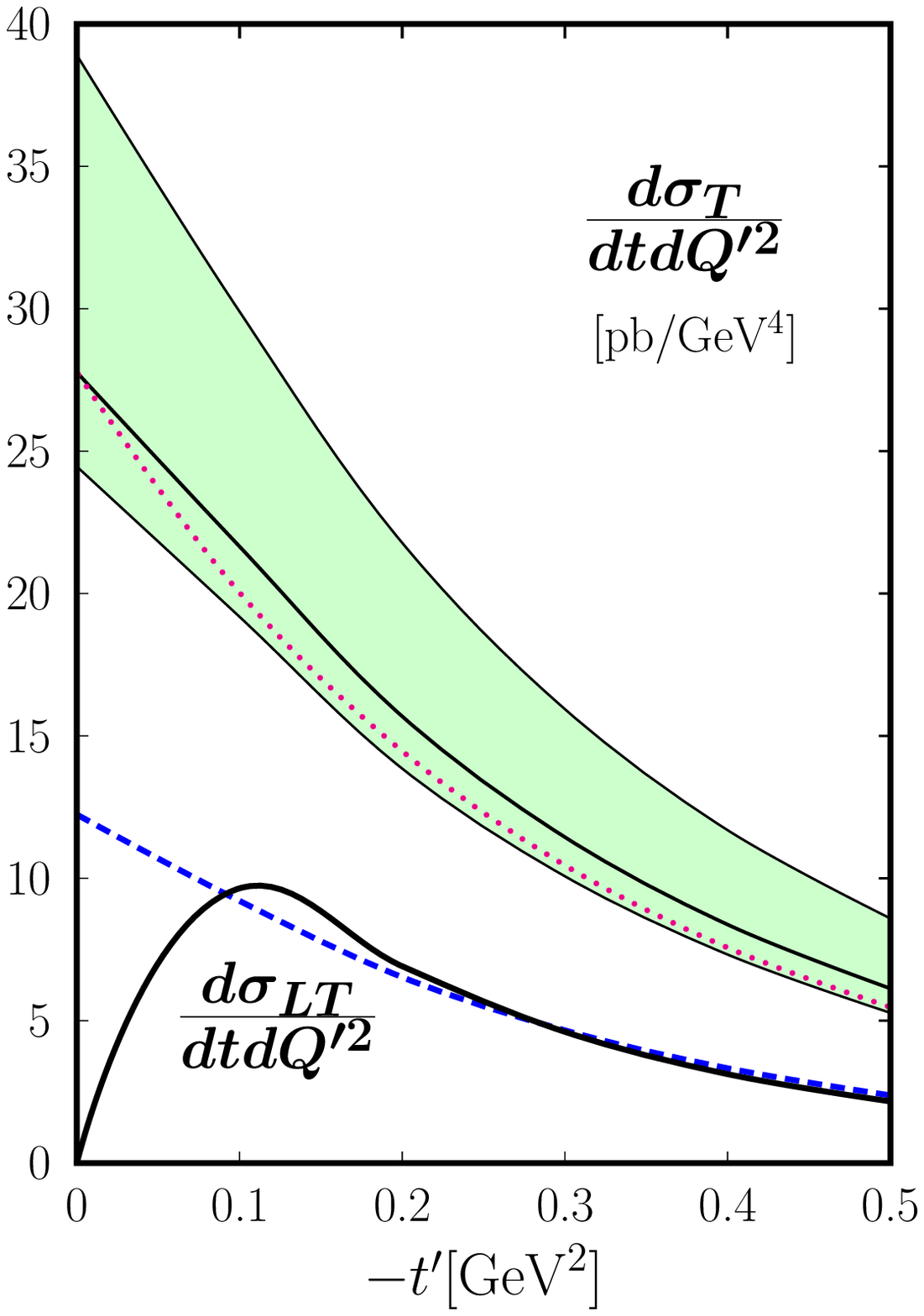}
\hspace*{0.06\tw}
\includegraphics[width=0.34\tw]{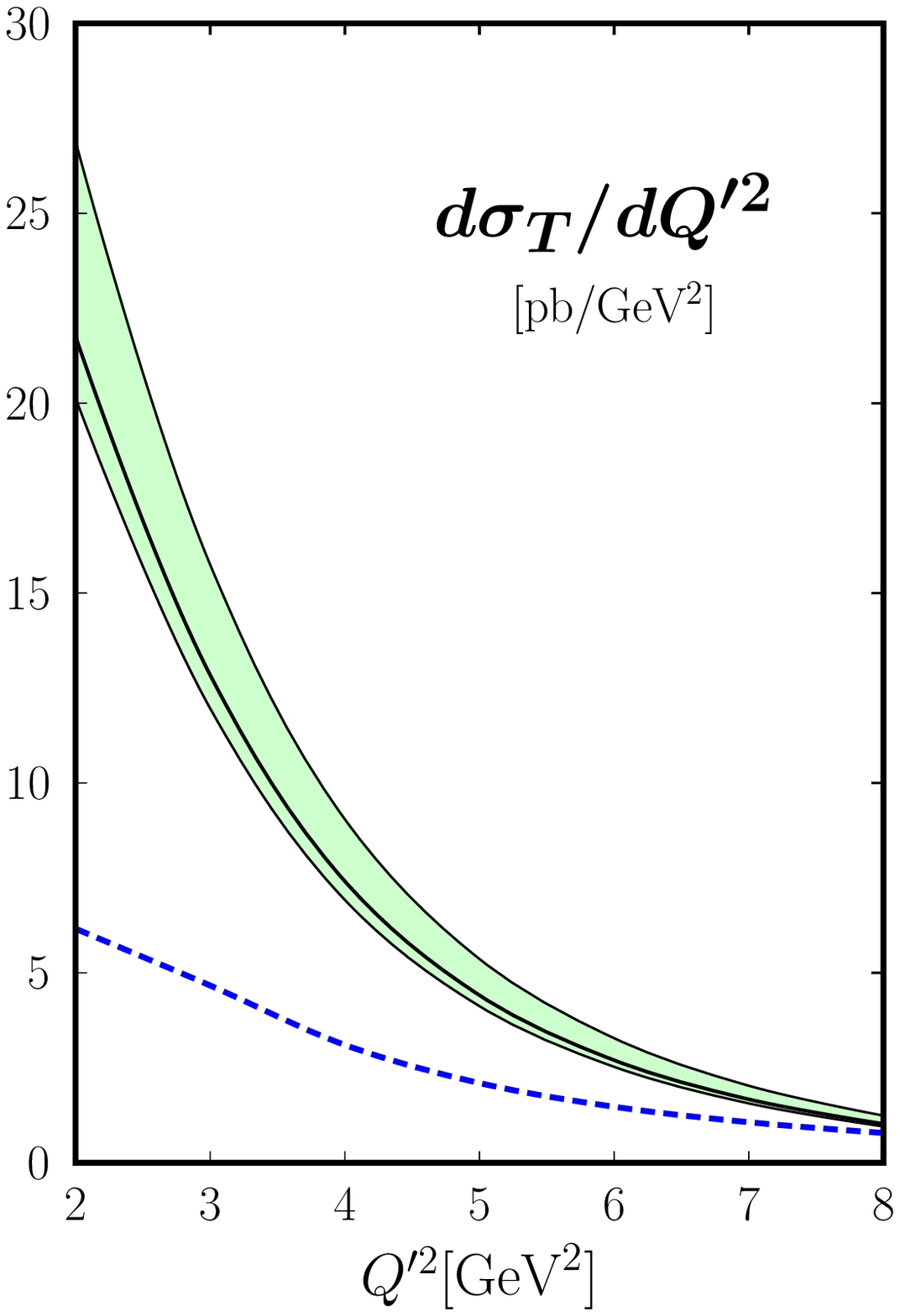}
\caption{The transverse cross sections $d\sigma_T/dtdQ^{\prime 2}$ (left) at 
$Q^{\prime 2}=4\,\gev^2$ versus $t'$  and $d\sigma_T/dQ^{\prime 2}$ (right) versus 
$Q^{\prime 2}$. The thin solid dashed lines with error bands represent the full 
result at $s=20\,\gev^2$, the thick dashed ones those at $30\,\gev^2$ while
the dotted line is the contribution from $H_T$. The thick solid line represents 
the longitudinal-transverse interference cross section.} 
\label{fig:sigmaT}
\end{figure}

The transverse cross section is shown in Fig.\ \ref{fig:sigmaT}.
It is substantially smaller than the longitudinal cross section but much larger 
than the leading-twist result. The uncertainty of our predictions is 
rather large and asymmetric due to the asymmetric error of $\mu_\pi$. 
The transverse cross section can be decomposed as (cf.\ \req{eq:partial} and 
\req{eq:amplitudes})~\footnote{
The $\bar{E}_T$ ($\pi$) term behaves as behaves as a natural (unnatural) parity 
exchange while the $H_T$ has no specific parity behavior \ci{GK5,GK6}.}
\be
\frac{d\sigma_T}{dtdQ^{\prime 2}} \= \kappa \big[ |{\cal M}_{--,0+}|^2 
           + 2|{\cal M}_{++,0+}(\pi)|^2
                               + 2|{\cal M}_{++,0+}(\bar{E}_T)|^2\big]\,.
\label{eq:transverse-cs}
\ee
The first term in \req{eq:transverse-cs}, being related to the GPD $H_T$, 
is displayed in Fig.\ \ref{fig:sigmaT} separately; it dominates this cross section. 
The second term, the pion-pole contribution, is rather small; it  
generates the little difference between the contribtuion from $H_T$ and the full result 
for $d\sigma_T$. The contribution from $\bar{E}_T$ is tiny. 

The longitudinal-transverse interference cross section is also shown in 
Fig.\ \ref{fig:sigmaT}. The width of its error band is about a half of that of 
the transverse cross section. $d\sigma_{LT}$ can be written as
\be
\frac{d\sigma_{LT}}{dtdQ^{\prime 2}}\= \kappa \operatorname{Re}\big[
       2{\cal M}^*_{0+,0+}{\cal M}_{++0+}(\pi) - {\cal M}^*_{0-,0+}{\cal M}_{--,0+}\big]\,.
\ee
Both the terms significantly contribute to $d\sigma_{LT}$.
The transverse-transverse interference cross section is given by 
\be
\frac{d\sigma_{TT}}{dtdQ^{\prime 2}}\=\kappa \big[ |{\cal M}_{++,0+}(\bar{E}_T)|^2
                                           -  |{\cal M}_{++,0+}(\pi)|^2\big]\,.
\ee
This cross section is very small. For instance, at $Q^{\prime 2}=4\,\gev^2$ and
$s=20\,\gev^2$ it is less than $\approx 0.3\,{\rm pb}/\gev^4$.

The cross sections decrease with growing $s$. As an example we show results
at $s=30\,\gev^2$ in the plots. At, say, $s\approx 360\,\gev^2$ as is available 
from the pion beam at CERN, the longitudinal cross section is about 
$30\,{\rm fb}/\gev^2$ at $Q^{\prime 2}=4\,\gev^2$. This is likely too small to be measured.

\section{Conclusions}
We calculated the partial cross sections for the exclusive Drell-Yan process,
$\pi^-p\to l^-l^+n$, within the handbag approach. In contrast to a previous
study of this process \ci{berger01} we treat the pion
pole as an OPE term and take into account transversity GPDs. The parametrizations
of the GPDs $\widetilde H$, $H_T$ and $\bar{E}_T$ as well as the values of other 
parameters appearing in the present calculation are taken from previous work 
\ci{GK5,GK6,DK13}. The generalization of our approach to $K^-p\to l^-l^+\Lambda$
is straightforward. 

Future data on $\pi^-p\to l^-l^+n$ measured at J-PARC may allow for a test of
factorization of the process amplitudes in hard subprocesses and soft GPDs. In
contrast to pion leptoproduction where there is a rigorous proof for factorization
of the amplitudes for longitudinally polarized photons, factorization of the
exclusive Drell-Yan process is an assumption although it seems plausible that
the factorization arguments also hold for time-like photons. However,  
Qiu \ci{qiu15} conjectured that factorization may be broken for the exclusive  
Drell-Yan process. If however factorization holds to a sufficient degree
of accuracy future data on the exclusive Drell-Yan process may improve our
knowledge of the GPDs.

The exclusive Drell-Yan process also offers the opportunity to check the 
dependence of the $\pi\pi\gamma$ vertex on the pion virtuality by comparing data 
on the time-like form factor measured in $l^+l^-\to\pi^+\pi^-$ with 
parametrizations of $\pi^-\pi^{+*}\to l^-l^+$ as part of the Drell-Yan analysis. 
The extraction of the space-like form factor from $lp\to l\pi^+n$ data may 
benefit from that check.

{\it Acknowledgements:} 
One of us (P.K.) likes to thank Markus Diehl and Oleg Teryaev for useful discussions
and remarks. The work is supported in part by the Heisenberg-Landau program and by 
the BMBF, contract number 05P12WRFTE.

\vskip 10mm

\end{document}